\documentstyle[12pt]{article}

\def\sub#1{\raisebox{-0.2ex}{\small#1}}
\begin{document}

\baselineskip=20pt plus 2pt
\hfill{NCKU-HEP/95-02}

\hfill{17 May 1995}
\begin{center}
\vspace{5mm}
{\large \bf
On the mass correction of heavy meson effective theory
\footnote{Research supported in part by the National Science
Council,The Republic of China.}\\}
\vspace{15mm}
Tsung-Wen Yeh and Chien-er Lee \vspace{5mm}

Department of Physics \\
National Cheng Kung University \\
Tainan, Taiwan 701, Republic of China \\
\end{center}
\vspace{15mm}
\begin{center}
{\bf ABSTRACT}
\end{center}
We derive the mass correction lagrangian of the heavy meson effective theory by
using the projection operator method. The next leading order of the mass
correction and  the first order of the chiral expansion are given
explicitely.We also consider the mass correction weak lagrangian of the heavy
mesons. Finally, we give the $\it{D}^{*+}\to \it{D}^o \pi^{+}$ decay
amplitude as an application.\\
\sub {PACS.11.30Rd -Chiral Symmetries}\\
\sub {PACS.13.20Fe -Chiral Lagrangian}
\newpage
The heavy quark physics have gotten much progress in recent years. One reason
comes from the discovery of the heavy quark symmetry of {\bf QCD} by N.
Isgur and M. B. Wise in {\bf 1989}[1]. This new symmetry appears in the
limit of the infinite mass of the heavy quark , and can be used to relate
 different matrix  elements of the heavy quark weak current. In {\bf 1990} ,
H. Georgi developed a heavy quark effective theory({\bf HQET}) to describe
 the above symmetry[2] for one heavy quark interacting with soft gluon by
assuming the velocity superselection rule.\\

In {\bf 1991}, many authors continued the program of {\bf HQET} to
construct the heavy hadron effective theory ({\bf HHET}) to combine
 heavy quark symmetry and chiral symmetry together[3]. As a member of the
 {\bf HHET}, the heavy meson effective theory ({\bf HMET}) considers
 one heavy meson interacting with the soft pseudoscalar pions[3].\\

Since the leading term in the lagrangian of {\bf HMET} only holds in the
infinite limit of the heavy quark mass , the mass correction effects need to be
taken into account. The $1/m_{Q}$ correction has been studied by H. Y. Cheng,
et al with interpolating field method and by N. Kitazawa and T. Kurimoto with
velocity reparametrization invariant method[4]. Here we apply the projection
operator method to derive the mass correction terms for {\bf HMET} up to
$1/m^{2}_{Q}$ order.\\

In this paper we construct the heavy meson effective lagrangian up to the
${\it O}(1/m^{2}_Q)$ corrections in $1/m_Q$ expansion and
${\it O}(p^2)$ in the chiral expansion. The lagrangian contains
nineteen parameters(with six new couplings, three new mass splitting parameters
between heavy pseudoscalar and vector mesons)necessary to be determined by
experiments.The light vector mesons are introduced via the hidden local
symmetry  method. The decay width of the
process ${\it D}^{*+}\to {\it D}^o \pi^+$ is calculated
with the $1/m^{2}_Q$ correction. Our notations are refered
to [4].\\

We denote the compact heavy meson field with large but finite mass by
${\it M}_v(x)$, which is defined as\\
\medskip
\begin{eqnarray}
{\it M}_v(x) &=& e^{im_{M}v \cdot x} {\it M}_{Q}(x) \nonumber\\
             &=& e^{im_{M}v \cdot x} \left( {\frac{i\not\!{d} + m_{M}}{2m_Q}}
                   \right) \left(- {\it P}_{Q} \gamma_5 +
                   {\it P}^{*\mu}_{Q}\gamma_{\mu}\right)(x) \nonumber\\
             &=& \left({\frac{i\not\!{d}+m_M+m_M\not\!{v}}{2m_Q}}\right)
                 \left(- {\widehat{\it P}}_{v} {\gamma_5}
                +{\widehat{\it P}}^{*\mu}_{v} \gamma_{\mu}\right)(x)
\nonumber\\
             &=& \left( {\frac{1+\not\!{v}}2} + {\frac{i\not\!{d}}{2m_Q}} +
{\frac{\Lambda}{m_Q}}{\frac{1+\not\!{v}}2 } \right) \left(-{\widehat{\it
P}}_{v} \gamma_5 +{\widehat{\it P}}^{*\mu}_{v} \gamma_{\mu}\right)(x),
\end{eqnarray}

\medskip
\noindent
where we have defined $m_M = m_Q +\Lambda$ and ${\it M}_{Q}(x)$ is the
compact heavy meson field defined in [3]. The well-known field
 ${\it H}_{v}(x)$ is defined as the infinite mass limit of
 ${\it M}_{v}(x)$

\begin{equation}
{\it H}_{v}(x)=\mathop{\lim_{m_{Q}\small{\to}\infty}}{\it M}_{v}(x)
\end{equation}

\noindent
and the relevant ingredients of ${\it H}_{v}(x)$ are also taken as the same
limit of those of ${\it M}_{v}(x)$

\begin{eqnarray}
{\it P}_{v}(x)&=&\mathop{\lim_{m_{Q}\small{\to}\infty}}
{\widehat{\it P}}_{v}(x) \nonumber\\
{\it P}^{*\mu}_{v}(x)&=&\mathop{\rm{\lim}}_{m_{Q}\small{\to}\infty}
{\widehat{\it P}}^{*\mu}_{v}(x).
\end{eqnarray}

 We take the pseudoscalar effective field ${\it P}_{v}(x)$ as example of
application of the projection operator method to derive its mass correction
 form. Transforming (1) to its momentum space form and taking the pseudoscalar
 part , we then have

\begin{eqnarray}
\left( {\frac{1+\not\!{v}}2} + {\frac{\not\!{k}}{2m_Q}} +
{\frac{\Lambda}{m_Q}}{\frac{1+\not\!{v}}2 } \right) {\widehat{\it P}}_{v}
&=&\left( {\frac{1+\not\!{v}}2} + {\frac{\not\!{k}}{2m_Q}}\right){\widehat{\it
P}}_{v} +
\left( {\frac{\Lambda}{m_Q}}{\frac{1+\not\!{v}}2 } \right){\widehat{\it
P}}_{v}
\nonumber\\ &\equiv&{\widehat{\it P}}_{v,Q} +{\widehat{\it P}}_{v,\bar{q}},
\end{eqnarray}

\noindent
where the ${\widehat{\it P}}_{v,Q} $ is the projected heavy degree of freedom
 and ${\widehat{\it P}}_{v,\bar{q}} $ is the projected light degree of
freedom.
 Such a definition is similar to the Bargman-Wigner wave function in the
 Bathe-Salpeter approach found by F.Hussain et al in[6] but here they are the
 field operators.The residual momentum $k$ comes from the momentum fluctuation
 of the heavy degree of freedom and thus is consistent with that definition of
 {\bf HQET}. To arrive at the mass expansion formalism, we need to further
 separate ${\widehat{\it P}}_{v,Q} $ into $(1+\not\!{v})/2$ and
$(1-\not\!{v})/2$ parts as

\begin{equation}
{\widehat{\beta}}_{v,Q}\equiv\left( {\frac{1+\not\!{v}}2}\right){\widehat{\it
P}}_{v,Q}
\,;\,{\widehat{\chi}}_{v,Q}\equiv\left(
{\frac{1-\not\!{v}}2}\right){\widehat{\it P}}_{v,Q}.\end{equation}

\noindent
We make an assumption that the heavy degree of freedom is on shell. Then the
 projection operator $((1+\not\!{v})/2 + {\not\!{k}}/{2m_Q})$ becomes
the energy projection operator(or, the spin projector) of the heavy quark.
There will be the following constrain on the residual momentum $k$

$$ v\cdot k=-{\frac{k^2}{2m_Q}}.$$

\noindent

Applying $((1+\not\!{v})/2 + {\not\!{k}}/{2m_Q})$ on ${\widehat{\it
P}}_{v,Q}$ , then we have

\begin{equation}
{\frac{1-\not\!{v}}2}{\widehat{\it P}}_{v,Q}={\frac{\not\!{k}}{2m_Q}}
{\widehat{\it P}}_{v,Q},
\end{equation}

\noindent
and

\begin{eqnarray}
{\widehat{\chi}}_{v,Q}&=&
{\frac{\not\!{k}}{2m_Q}}\left({\widehat{\beta}}_{v,Q}+
                      {\widehat{\chi}}_{v,Q}\right) \nonumber\\
                  &=&
{\frac{\not\!{k}}{2m_Q-{\not\!{k}}}}{\widehat{\beta}}_{v,Q},
\end{eqnarray}

\noindent
These then give

\begin{equation}
{\widehat{\it P}}_{v,Q}={\frac{2m_Q}{2m_Q-{\not\!{k}}}}{\widehat{\beta}}_{v,Q}
\end{equation}

  Therefore we only need to determine ${\widehat{\beta}}_{v,Q}$. By the
follwing ansatz

\begin{eqnarray}
{\widehat{\beta}}_{v,Q}=\left( 1+{\widehat{\omega}}_{v,Q}\right){\it P}_{v,h}
\nonumber\\
{\overline{\widehat{\omega}}}_{v,Q}={\gamma}^{o}{\widehat{\omega}}^{\dagger}_{v
, Q } {\gamma}^{o}={\widehat{\omega}}_{v,Q},
\end{eqnarray}

\noindent
and noting that

\begin{equation}
{\Lambda}^{+}_{Q}={\mathop{\sum}_{r}}{\widehat{\it P}}_{r,v,Q}
{\overline{\widehat{\it P}}}_{r,v,Q} \,\, , \,\,
{\Lambda}^{+}_{h}={\mathop{\sum}_{r}}{\it P}_{r,v,h}
{\overline{\it P}}_{r,v,h},
\end{equation}

\noindent
with ${\Lambda}^{+}_{Q}=((1+\not\!{v})/2 + {\not\!{k}}/{2m_Q})$ ,
${\Lambda}^{+}_{h}=(1+\not\!{v})/2 $ ,
 ${\it P}_{r,v,h}=([(1+\not\!{v})/2]{\it P}_{v})_r$, we obtain

\begin{equation}
\left(
1+{\widehat{\omega}}_{v,Q}\right){\Lambda}^{+}_{h}\left(
1+{\widehat{\omega}}_{v , Q }
\right) =\left( 1-{\frac{\not\!{k}}{2m_Q}}\right){\Lambda}^{+}_{Q}\left(
1-{\frac{\not\!{k}}{2m_Q}}\right).
 \end{equation}

\noindent
(11) implies the following quadratical equation

\begin{equation}
{\widehat{\omega}}^{2}_{v,Q}+2{\widehat{\omega}}_{v,Q}-{\widehat{\it
T}}_{v,Q}=0,
\end{equation}

\noindent
where ${\widehat{\it T}}_{v,Q}$ is given by

\begin{eqnarray}
{\widehat{\it T}}_{v,Q} & = &
\left( 1-{\frac{\not\!{k}}{2m_Q}}\right){\Lambda}^{+}_{Q}\left(
1-{\frac{\not\!{k}}{2m_Q}}\right) -{\Lambda}^{+}_{h} \nonumber \\
& = & -\left( {\frac{\not\!{k}}{2m_Q}}\right)^{2}{\Lambda}^{+}_{h},
\end{eqnarray}
\noindent
and is of ${\it O}(m^{-2}_{Q})$. Since ${\widehat{\omega}}_{v,Q}\ll 1$, the
relevant solution is

\begin{eqnarray}
{\widehat{\omega}}_{v,Q}& = & -1 + \left( 1 + {\widehat{\it
T}}_{v,Q}\right)^{1/2} \nonumber \\
 &=& {\frac{1}2} {\widehat{\it T}}_{v,Q}-
{\frac{1}8}{\widehat{\it T}}^{2}_{v,Q}+{\frac{1}{16}}
{\widehat{\it T}}{3}_{v,Q}-{\frac{5}{128}}
{\widehat{\it T}}^{4}_{v,Q}+\ldots.
\end{eqnarray}

We note that the ${\widehat{\omega}}_{v,Q}$ in (14) agrees to the ansatz
(9). By using

\begin{equation}
\left({\widehat{\it T}}_{v,Q}\right)^{n}{\it P}_{v,h}
=(-)^{n}\left( {\frac{\not\!{k}}{2m_Q}}\right)^{2n}{\it P}_{v,h},
\end{equation}

\noindent
we obtain

\begin{equation}
{\widehat{\it P}}_{v,Q} =
 \left( {\frac{1+{\not\!{k}}/{2m_Q}}{1-{\not\!{k}}/{2m_Q}}}\right)^{1/2}
{\it P}_{v,h}.
\end{equation}

\noindent
To determine ${\widehat{\it P}}_{v,{\bar{q}}}$ , we can use the following
useful relations

\begin{equation}
\left[{\frac{1+{\not\!{v}}}{2}}\, , \,
{\sqrt{\frac{1+{\not\!{k}}/{2m_Q}}{1-{\not\!{k}}/{2m_Q}}}}\right]
=-{\frac{\not\!{k}}{2m_Q}}
{\sqrt{\frac{1+{\not\!{k}}/{2m_Q}}{1-{\not\!{k}}/{2m_Q}}}}
\end{equation}

\begin{equation}
{\sqrt{\frac{1+{\not\!{k}}/{2m_Q}}{1-{\not\!{k}}/{2m_Q}}}}
\left( {\frac{1+{\not\!{v}}}{2}}\right){\it P}_{v,h}
=\left( {\frac{1+{\not\!{v}}}{2}}+{\frac{\not\!{k}}{2m_Q}}\right)
{\sqrt{\frac{1+{\not\!{k}}/{2m_Q}}{1-{\not\!{k}}/{2m_Q}}}}{\it P}_{v,h}.
\end{equation}

\noindent
Then (4) and (18) imply that ${\widehat{\it P}}_{v}$ relates to ${\it
P}_{v,h}$ as

\begin{equation}
{\widehat{\it P}}_{v}=
{\sqrt{\frac{1+{\not\!{k}}/{2m_Q}}{1-{\not\!{k}}/{2m_Q}}}}{\it P}_{v,h}.
\end{equation}

\noindent
Thus ${\widehat{\it P}}_{v,{\bar{q}}}$ relates to ${\it P}_{v,h} $ as

\begin{equation}
{\widehat{\it P}}_{v,{\bar{q}}}={\frac{\Lambda}{m_Q}}
\left( {\frac{1+{\not\!{v}}}{2}} \right)
{\sqrt{\frac{1+{\not\!{k}}/{2m_Q}}{1-{\not\!{k}}/{2m_Q}}}}{\it P}_{v,h}.
\end{equation}

     We can do the same procedure for the vector meson field ${\widehat{\it
P}}^{*\mu}_{v}$ and combine all results to arrive at

\begin{equation}
{\it M}_{v}=\left( 1+{\frac{\Lambda}{m_Q}}{\frac{1+{\not\!{v}}}{2}} \right)
{\sqrt{\frac{1+{\not\!{k}}/{2m_Q}}{1-{\not\!{k}}/{2m_Q}}}}{\it H}_{v}.
\end{equation}

\noindent
In the above , we have arrived at the complet mass correction form for compact
heavy meson effective field. It should be noted that there is no any ambiguity
in this mass correction scheme. As pointed out by H.Y.Cheng, et al in[4] ,
there exists some ambiguities in the mass expansion via velocity
reparametrization invariant method(the reason see the original paper).
 And such a problem has been solved partly
by N.Kitazawa and T.Kurimoto by using trial and error in their derivation of
mass correction lagarangian[4]. The reason about our approach to not have
such a ambiguity is coming from that we directly solve the mass correction
equation (4). \\

It needs a short comment about the meaning of the mass correction form (21).
To our opinion, the key ingredient of the heavy quark mass expansion should be
the field mass corrction form for the infinite mass limit field. This is
because that:(1) from the renormalization group point of view , we need an
effective field theory which is continuously matching down to from the complete
theory along the relevant renormalization group ({\bf RG}) flows. In this
intermediate scale region between effective and complete theories, {\bf RG}
flows are affected mostly from the heavy quark part. The mass expansion at some
order should be complete under {\bf RG} operation. This is the renormalizable
problem of the heavy quark mass exapnsion. The mass expansion scheme is used to
build the basis for the {\bf RG} operation. (2) Any heavy quark involved
in the green function under {\bf RG} operation always needs one or more heavy
quark fields for the calculation of its renormalization constant , both in
effective and complete theories. (3)From the field theory point of view the
difference between the complete and effective theory is the existence of an
exotic symmetry group for effective theoy. The mass term will certainly break
such symmetry. So ,the mass expansion can be seen as an transformation between
effective field and original field. The conclusion is that from theoretical
and practical view point the field mass correction form is most appropriate for
needs.\\

The following we can write down the effective lagrangian which is
invariant under parity transformation and charge conjugation up to
${\it O}(1/m^{2}_{Q})$ by using (21). All the possible terms are

\begin{eqnarray}
{\it L}& = & {{\it L}_{particle}}+{{\it L}_{antiparticle}} \nonumber \\
        & = &- {\mathop{\sum}_{v}}{tr}\{ {\overline{\it H}}_{v}
        {v \cdot iD}{\it H}_{v} \} - {\mathop{\sum}_{v}}
        {tr}\{ {\overline{\it H}}_{v} {\frac{(i{\not\!{D}})^2}{2M}}
        {\it H}_{v} \} \nonumber \\
 & & - {\mathop{\sum}_{v}}
        {tr}\{ {\overline{\it H}}_{v} {\frac{(i{\not\!{D}})^3}{2M^2}}
        {\it H}_{v} \} + { h.c.}\nonumber \\
& & + {\Lambda}{\mathop{\sum}_{v}}{tr}\{ {\overline{\it H}}_{v}
     {\it H}_{v} \}
- {\Lambda}{\mathop{\sum}_{v}}{tr}\{ {\overline{\it H}}_{v}
{\frac{(v\cdot iD)}{2M}}{\it H}_{v} \} + { h.c.} \nonumber \\
& & +{\Lambda}{\mathop{\sum}_{v}}{tr}\{ {\overline{\it H}}_{v}
{\frac{(i{\not\!{D}})^2}{2M^2}}{\it H}_{v} \} + {\kappa}_{1}
{\Lambda}{\mathop{\sum}_{v}}{tr}\{ {\overline{\it H}}_{v}
{\frac{\Lambda}{2M}}{\it H}_{v} \}   \nonumber \\
& & +{\kappa}_{2}{\Lambda}{\mathop{\sum}_{v}}{tr}\{ {\overline{\it H}}_{v}
{\frac{\Lambda}{2M}}{\sigma}_{\mu\nu}{\it H}_{v}{\sigma}^{\mu\nu}\}
-{\kappa}_{3}{\Lambda}{\mathop{\sum}_{v}}{tr}\{ {\overline{\it H}}_{v}
{\frac{{\Lambda}^2}{M^2}}{\it H}_{v} \} \nonumber\\
& & -{\kappa}_{4}{\Lambda}{\mathop{\sum}_{v}}{tr}\{ {\overline{\it H}}_{v}
{\frac{{\Lambda}^2}{2M^2}}{\sigma}_{\mu\nu}{\it H}_{v}{\sigma}^{\mu\nu}\}
+{\gamma}{\mathop{\sum}_{v}}{tr}\{ {\overline{\it H}}_{v}{\it H}_{v}
{v \cdot {\widehat{\alpha}}_{\parallel}} \} \nonumber\\
&& + {\gamma}{\mathop{\sum}_{v}}{tr}\{ {\overline{\it H}}_{v} {\it H}_{v}
{v \cdot {\widehat{\alpha}}_{\parallel}}\}
+ {\gamma}{\mathop{\sum}_{v}}{tr}\{ {\overline{\it H}}_{v}
{\frac{iD^{\mu}}{2M}}{\it H}_{v}{\widehat{\alpha}}_{\parallel\mu}\}+{h.c.}
\nonumber\\
&& + {\gamma}{\mathop{\sum}_{v}}{tr}\{ {\overline{\it H}}_{v}
{\frac{v \cdot {iD}}{2M}}{\it H}_{v}
{v \cdot {\widehat{\alpha}}_{\parallel}}\}+{h.c.} \nonumber\\
&&+{\gamma}_{1}{\mathop{\sum}_{v}}{tr}\{ {\overline{\it H}}_{v}
{\frac{\Lambda}{2M}}{\it H}_{v}{v \cdot {\widehat{\alpha}}_{\parallel}}\}
+{\gamma}_{2}{\mathop{\sum}_{v}}{tr} {\overline{\it H}}_{v}
{\frac{\Lambda}{2M}}{\sigma}_{\mu\nu}{\it H}{\sigma}^{\mu\nu}
{v \cdot {\widehat{\alpha}}_{\parallel}}\} \nonumber\\
&& - {\gamma}_{3}{\Lambda}{\mathop{\sum}_{v}}{tr}\{ {\overline{\ H}}_{v}
{\frac{5{\Lambda}}{2M^2}}{\it H}_{v} {v \cdot {\widehat{\alpha}}_{\parallel}}
\} + {\gamma}_{4}{\Lambda}{\mathop{\sum}_{v}}{tr}\{ {\overline{\it H}}_{v}
{\frac{2iD^{\mu}}{2M^2}}{\it H}_{v}{\widehat{\alpha}}_{\parallel\mu}\}+{h.c.}
\nonumber\\
&& - {\gamma}_{5}{\Lambda}{\mathop{\sum}_{v}}{tr}\{ {\overline{\ H}}_{v}
{\frac{\Lambda}{2M^2}}{\sigma}_{\mu\nu}{\it H}{\sigma}^{\mu\nu}
{v \cdot {\widehat{\alpha}}_{\parallel}}\}\nonumber\\
&& + {\gamma}{\mathop{\sum}_{v}}{tr}\{ {\overline{\it H}}_{v}
{\frac{1}{2M^2}}\{ \left[ {(iD)^2}v^{\mu}+({v \cdot iD}) {iD^{\mu}}
+ {iD^{\mu}}({v \cdot iD}) \right] \} {\it
H}_{v}{\widehat{\alpha}}_{\parallel\mu} \} \nonumber\\
&& + {\gamma}{\mathop{\sum}_{v}}{tr}\{ {\overline{\it H}}_{v}
{\frac{1}{2M^2}}{i{\epsilon}^{\mu\nu\rho\sigma}}{iD_{\mu}}{iD_{\nu}}
{\it H}_{v}{\gamma}_{5}{\widehat{\alpha}}_{\parallel\rho}{\gamma}_{\sigma}\}
+{h.c.}\nonumber\\
&&- {\lambda}{\mathop{\sum}_{v}}{tr}\{ {\overline{\ H}}_{v}{\ H}_{v}
{\gamma}_{5}{\not\!{\alpha}}_{\perp}\}
+ {\lambda}{\mathop{\sum}_{v}}{tr}\{ {\overline{\it H}}_{v}
{\frac{v \cdot iD}{2M}}{\it H}_{v}{\gamma}_{5}{\not\!{\alpha}}_{\perp}\}
+{h.c.}\nonumber\\
&& - {\lambda}{\mathop{\sum}_{v}}{tr}\{ {\overline{\it H}}_{v}
{\frac{{\epsilon}^{\mu\nu\rho\sigma}}{4M}}{iD_{\rho}}{\it H}_{v}
{\sigma}_{\mu\nu}{\alpha}_{\perp\sigma}\} + {h.c.}\nonumber\\
&& - {\lambda}_{1}{\mathop{\sum}_{v}}{tr}\{ {\overline{\it H}}_{v}
{\frac{\Lambda}{2M}}{\it H}_{v}{\gamma}_{5}{\not\!{\alpha}}_{\perp}\}
 - {\lambda}_{2}{\mathop{\sum}_{v}}{tr}\{ {\overline{\it H}}_{v}
{\frac{\Lambda}{2M}}{\gamma}_{5}{\gamma}_{\rho}{\it H}_{v}
{\alpha}_{\perp}^{\rho}\} \nonumber\\
&& + {\lambda}_{3}{\Lambda}{\mathop{\sum}_{v}}{tr}\{ {\overline{\it H}}_{v}
{\frac{5\Lambda}{2M^2}}{\it H}_{v}{\gamma}_{5}{\not\!{\alpha}}_{\perp}\}
+ {\lambda}_{4}{\Lambda}{\mathop{\sum}_{v}}{tr}\{ {\overline{\it H}}_{v}
{\frac{5\Lambda}{2M^2}}{\gamma}_{5}{\gamma}_{\rho}{\it H}_{v}
{\alpha}_{\perp}^{\rho}\} \nonumber\\
&& - {\lambda}_{5}{\Lambda}{\mathop{\sum}_{v}}{tr}\{ {\overline{\it H}}_{v}
{\frac{2v \cdot iD}{M^2}}{\it H}_{v}{\gamma}_{5}{\not\!{\alpha}}_{\perp}\}
+{h.c.}\nonumber\\
&& - {\lambda}_{6}{\Lambda}{\mathop{\sum}_{v}}{tr}\{ {\overline{\it H}}_{v}
{\frac{5{\Lambda}{\epsilon}^{\mu\nu\rho\sigma}}{4M^2}}{iD_{\rho}}{\it H}_{v}
{\sigma}_{\mu\nu}{\alpha}_{\perp\sigma}\} + {h.c.}\nonumber\\
&& + {\lambda}{\mathop{\sum}_{v}}{tr}\{ {\overline{\it H}}_{v}
{\frac{1}{2M^2}}\{ \left[ {(iD)^2}{\gamma}^{\mu}+({i{\not\!D}}) {iD^{\mu}}
+ {iD^{\mu}}({i{\not\!D}}) \right] \} {\it
H}_{v}{\gamma}_{5}{\alpha}_{\perp\mu} \} \nonumber\\
&& + {\lambda}{\mathop{\sum}_{v}}{tr}\{ {\overline{\it H}}_{v}
{\frac{1}{2M^2}}{i{\epsilon}^{\mu\nu\rho\sigma}}{iD_{\mu}}{iD_{\nu}}
{\it H}_{v}{\alpha}_{\perp\rho}{\gamma}_{\sigma}\}
+{h.c.}\nonumber\\
&&+{\it L}_{antiparticle},
\end{eqnarray}
\noindent
where $1/M=diag(1/m_c,1/m_b)$,$1/M^{2}=diag(1/m_c^2,1/m_b^2)$.
The anti-particale type lagrangian ${\it L}_{antiparticle}$
can be obtained by substituting ${\it H}_{v}{\to}{\it H}_{v}^{-}$, and
$v \to -v$. For comparison , we also write down the pseudoscalar
 meson masses $m_P$ and the vector meson masses $m_V$ to be expanded as

$$
{m_V}^{2} = {m_Q}^{2} \{ 1+2{\frac{\Lambda}{m_Q}}+{\kappa}_{1}
{\frac{{\Lambda}^2}{m_Q^2}}+6{\kappa}_{2}{\frac{{\Lambda}^3}{m_Q^2}}
-2{\kappa}_{3}{\frac{{\Lambda}^3}{m_Q^3}}-6{\kappa}_{4}
{\frac{{\Lambda}^3}{m_Q^3}}\} $$
\begin{equation}
 {m_P}^{2} = {m_Q}^{2} \{ 1+2{\frac{\Lambda}{m_Q}}+{\kappa}_{1}
{\frac{{\Lambda}^2}{m_Q^2}}-2{\kappa}_{2}{\frac{{\Lambda}^3}{m_Q^2}}
-2{\kappa}_{3}{\frac{{\Lambda}^3}{m_Q^3}}+2{\kappa}_{4}
{\frac{{\Lambda}^3}{m_Q^3}}\}.
\end{equation}
\noindent
We adope the effective field normalization convention of [4] as
$\sqrt{m_Q}$. The covariant derivative $iD_{\mu}$ is defined as
\begin{equation}
iD_{\mu}{\it H}_{v}=i{\partial}_{\mu}{\it H}_{v}(x)-{\it H}_{v}(x)
g_VV^a_{\mu}(x){\frac{\lambda^a}{2}}
\end{equation}

\noindent
to include the light vector meson fields $V^a_{\mu}$.\\

For the heavy-to-light weak lagrangian we can also obtain it in the
$1/m_Q$ expansion and in the chiral expansion as

\begin{eqnarray}
{\it L}_W &=&i \{ {\frac{F}{2}} tr\{ J{\xi}^{\dagger}
[1+{\frac{i{\not\!D}}{2m_Q}}
+{\frac{(i{\not\!D})^2}{8m_Q^2}}]{\it H}_{v} \} \nonumber\\
&&+a_1{\frac{\Lambda}{2m_Q}} tr\{ J{\xi}^{\dagger}{\it H}_{v} \}
+a_2{\frac{\Lambda}{2m_Q}} tr\{ J{\xi}^{\dagger}{\gamma}^{\rho}
{\it H}_{v}{\gamma}_{\rho} \} \nonumber \\
&& + a_3{\frac{\Lambda}{4m_Q^2}} tr\{ J{\xi}^{\dagger}{v \cdot iD}
{\it H}_{v} \}
+ a_4{\frac{\Lambda}{4m_Q^2}} tr\{ J{\xi}^{\dagger}{v \cdot iD}{\gamma}^{\rho}
{\it H}_{v}{\gamma}_{\rho} \} \nonumber \\
&& + a_5{\frac{\Lambda^2}{8m_Q^2}} tr\{ J{\xi}^{\dagger}{\it H}_{v} \}
+ a_6{\frac{\Lambda^2}{8m_Q^2}} tr\{ J{\xi}^{\dagger}{\gamma}^{\rho}
{\it H}_{v}{\gamma}_{\rho} \} \nonumber \\
&& + a_7{\frac{\Lambda}{8m_Q^2}} tr\{ J{\xi}^{\dagger}{\sigma}_{\mu\nu}
{\it H}_{v} {\sigma}^{\mu\nu}\}  \nonumber \\
&& + b_1 tr\{ J{\xi}^{\dagger}{\it
H}_{v}{\widehat{\not\!{\alpha}}}_{\parallel} \}
 + b_2 tr\{ J{\xi}^{\dagger}{\it
H}_{v}{v \cdot {\widehat{\alpha}}_{\parallel}} \} \nonumber\\
&& + c_1{\frac{1}{4m_Q}} tr\{ J{\xi}^{\dagger}{iD^{\mu}}
{\it H}_{v}{\widehat{\alpha}}_{\parallel\mu} \}
+ c_2{\frac{1}{2m_Q}} tr\{ J{\xi}^{\dagger}i{\not\!D}
{\it H}_{v}{\widehat{\not\!{\alpha}}}_{\parallel} \} \nonumber\\
&& + c_3{\frac{\Lambda}{2m_Q}} tr\{ J{\xi}^{\dagger}{\it H}_{v}
{\widehat{\not\!{\alpha}}}_{\parallel} \}
+ c_4 {\frac{1}{8m_Q}} tr\{ J{\xi}^{\dagger}(i{\not\!D})^2
{\it H}_{v}{\widehat{\not\!{\alpha}}}_{\parallel} \} \nonumber\\
&& + c_5{\frac{\Lambda}{4m_Q^2}} tr\{ J{\xi}^{\dagger}{v \cdot iD}
{\it H}_{v}{\widehat{\not\!{\alpha}}}_{\parallel} \}
+ c_6{\frac{\Lambda^2}{8m_Q^2}} tr\{ J{\xi}^{\dagger}{\it H}_{v}
{\widehat{\not\!{\alpha}}}_{\parallel} \} \nonumber\\
&& + c_7{\frac{\Lambda^2}{8m_Q^2}} tr\{ J{\xi}^{\dagger}{\sigma}_{\mu\nu}
{\it H}_{v}{\sigma}^{\mu\nu}
{\widehat{\not\!{\alpha}}}_{\parallel} \}
+ c_8{\frac{1}{2m_Q}} tr\{ J{\xi}^{\dagger}i{\not\!D}
{\it H}_{v}{v \cdot {\widehat{\alpha}}_{\parallel}} \} \nonumber\\
&& + c_9{\frac{\Lambda}{2m_Q}} tr\{ J{\xi}^{\dagger}{\it H}_{v}
{v \cdot {\widehat{\alpha}}_{\parallel}} \}
+ c_{10}{\frac{1}{8m_Q^2}} tr\{ J{\xi}^{\dagger}(i{\not\!D})^2
{\it H}_{v}{v \cdot {\widehat{\alpha}}_{\parallel}} \} \nonumber\\
&& + c_{11}{\frac{\Lambda}{4m_Q^2}} tr\{ J{\xi}^{\dagger}{v \cdot iD}
{\it H}_{v}{v \cdot {\widehat{\alpha}}_{\parallel}} \}
+ c_{12}{\frac{\Lambda}{4m_Q^2}} tr\{ J{\xi}^{\dagger}
{\it H}_{v}{v \cdot {\widehat{\alpha}}_{\parallel}} \} \nonumber\\
&& + c_{13}{\frac{\Lambda^2}{8m_Q^2}} tr\{ J{\xi}^{\dagger}{\sigma}_{\mu\nu}
{\it H}_{v}{\sigma}^{\mu\nu}
{v \cdot {\widehat{\alpha}}_{\parallel}} \} \},
\end{eqnarray}
\noindent
where $J\equiv J^{\mu}{\gamma}_{\mu}(1-{\gamma}_5)$
,is regarded as an external current and also invariant under parity
transformation and we use the convention  $J^{\mu\dagger}\equiv(Q{\bar
q})$,$Q=(c,b),{\bar q}=({\bar u} ,{\bar d},{\bar s} )$.
We can extract the decay constants of the heavy meson from the above weak
lagrangian

\begin{equation}
f_p = {\sqrt{\frac{2}{m_Q}}}\{
F+{\frac{\Lambda}{m_Q}}(a_1+a_2)+{\frac{\Lambda^2}{4m_Q^2}}(a_5+a_6-4a_7)\}
\end{equation}

\begin{equation}
f_V = {\sqrt{\frac{2}{m_Q}}}\{
F+{\frac{\Lambda}{m_Q}}(a_1-a_2)+{\frac{\Lambda^2}{4m_Q^2}}(a_5-a_6+12a_7)\}.
\end{equation}

We calculate the decay width of ${\it D}^{*+}\to {\it D}^o{\pi}^+$ as
the application of our formalism. The result is

\begin{eqnarray}
\lefteqn{{\Gamma}({\it D}^{*+}{\to} {{\it D}^o}{{\pi}^+})} \nonumber\\
& & {\approx}
{\frac{{\lambda^2}{m_c^2}{({E_{\pi}^2}-{m_{\pi}^2})^{1/2}}}
{12{\pi}{m_{D^*}^2}{f^2_{\pi }}}} \{ 1+
 {E_{\pi}}\left( {\frac{1}{m_c}}
+{\frac{5{\lambda_6}{\Lambda^2}}{m_c^2}}\right) \nonumber \\
& & + {\frac{({\lambda_1}-{\lambda_2}){\Lambda}}{{\lambda}{m_c}}}
-{\frac{5({\lambda_3}-{\lambda_4}){\Lambda^2}}{{\lambda}{m_c^2}}}
 \},
\end{eqnarray}
\noindent
where

$$
E_{\pi}={\frac{{m^2_{D^*}}-{m^2_D}+{m^2_{\pi}}}{2m_{D^*}}},
$$
\noindent
and we have dropped some double counting form factor terms.\\

In this paper , we have derived the mass correction of HMET by means
 of the projection operator method. We construct the heavy meson effective
 lagrangian up to ${\it O}(1/m_Q^2)$ in the $1/m_Q$ expansion
 and ${\it O}(p^2)$ in the chiral expansion.For the further applications we
leave them to our future publications.\\


\begin{thebibliography}{99}
\bibitem[1]{}N.Isgur and M.B.Wise,Phys.Lett.B 232(1989)113.
\bibitem[2]{}H.Georgi,Phys.Lett.B 240(1990)447.
\bibitem[3]{}E.Jenkins and A.V.Manohar,Phys.Lett.B 235(1991)558;
    G.Burdman %
    and J.F.Donoghue,Phys.Lett.B 280(1992)287;P.Cho and H.Georgi,
    Phys.Lett.B 296(1992)408;T-M.Yan, H-Y.Cheng, C-Y.Cheung, G-L.Lin,%
    Y.C.Lin,and H-L.Yu, Phys.Rev.D 46(1992)1148.%
\bibitem[4]{}H-Y.Cheng, C-Y.Cheung, G-L.Lin, Y-C.Lin, T-M.Yan and
    H-Y.Yu, ClNS93/1192;%
    N.Kitazawa and T.Kurimoto, Phys.Lett.B323(1994)65.%
\bibitem[5]{}Chien-er Lee and T-W Yeh,(unpublished)%
\bibitem[6]{}S.Balk, J.G.Korner, G.Thompson and F.Hussain, IC/91/397

\end{thebibliography}
\end{document}